\documentclass[12pt]{article}
\usepackage{graphicx}
\usepackage{epsfig,cite}
\usepackage{amssymb}
\usepackage{amsmath}
\usepackage{dsfont}
\usepackage{multirow}
\usepackage{color}
\usepackage{subfigure,amstext,alltt,setspace}
\usepackage{amsbsy}
\usepackage{comment}
\usepackage{fullpage}
\usepackage{array}
\usepackage{booktabs,multirow,tabularx}
\usepackage{hyperref}
\usepackage{slashed}
\usepackage{url}

\newcommand{\order}[1]{{\cal O}(#1)}

\begin{document}

\begin{flushright}
TIF-UNIMI-2015-12\\
IPPP/15/51 \\
DCPT/15/102 \\
% HT--15--?? \\
Cavendish-HEP-15/06 
\end{flushright}

\vspace*{1cm}

\begin{center}
 {\Large \bf{Higgs production in bottom-quark fusion\\ in a matched scheme}}
\end{center}

\vspace*{.7cm}

\begin{center}
 Stefano Forte$^{1}$, Davide Napoletano$^2$ and Maria Ubiali$^{3}$
\vspace*{.2cm}

\noindent
{\it
  $^1$ Tif Lab, Dipartimento di Fisica, Universit\`a di Milano and\\ 
INFN, Sezione di Milano,
  Via Celoria 16, I-20133 Milano, Italy\\
  $^2$ Institute for Particle Physics Phenomenology,\\
  Durham University, Durham DH1 3LE, UK\\
  $^3$ Cavendish Laboratory, University of Cambridge,\\
  J.J. Thomson Avenue, CB3 0HE, Cambridge UK\\}

\vspace*{3cm}

%\begin{center}
{\bf Abstract}
\end{center}

\noindent
We compute the total cross-section for Higgs boson production in
bottom-quark fusion using the 
so-called FONLL method for the matching of a scheme in which the $b$-quark is
treated as a massless parton to that in which it is treated as a
massive final-state particle.
We discuss the general framework for the application of the FONLL method
to this process, and then we present explicit expressions for the case
in which the next-to-next-to-leading-log five-flavor scheme result is
combined with the leading-order $\order{\alpha_s^2}$
four-flavor scheme computation. We compare our results in this case 
to the four-and five-flavor scheme computations, 
and to the so-called Santander matching.

\pagebreak

%\tableofcontents

%%%%%%%
%\section{Introduction}
In perturbative QCD, processes involving bottom quarks can be computed 
within different factorization schemes. One possibility is to use a
five-flavor, or massless, scheme, in which
the $b$-quark is treated as a massless parton. In this scheme,
collinear logarithms of $\mu_F^2/m_b^2$ (with $\mu_F$ the
factorization scale) are resummed through QCD
evolution equations, but corrections suppressed by powers of
$m_b^2/\mu_F^2$ are neglected. Alternatively, one may use 
a four-flavor, massive, or decoupling scheme, 
in which the $b$-quark is treated as a massive particle, which
decouples from evolution equations and the running of $\alpha_s$, but full
dependence on $m_b$ is retained. 
Generally, of course, results in the two scheme
may differ by a large amount: indeed, the leading-order predictions
for Higgs boson in bottom-quark fusion~\cite{Kunszt:1984ri,Dicus:1988cx,Barnett:1987jw,Dicus:1998hs}
may differ by up to one order of magnitude~\cite{Spira:1998wh}, though
the disagreement is reduced if the factorization and renormalization
scales are chosen to be smaller than $m_H$ (which may
well~\cite{Rainwater:2002hm,Plehn:2002vy,Maltoni:2003pn,Maltoni:2012pa,Ubiali:2014cva}
be more appropriate) and higher perturbative orders are included.

The five-flavor scheme is more accurate
for scales $\mu^2\gg m_b^2$, while the four-flavor scheme is more accurate close to
threshold,   though of course if the four-flavor computation is performed to
high enough order in perturbation theory it will reproduce the
five-flavor scheme result (the converse is not true, because mass
corrections are not included in the five-flavor scheme at any
perturbative order). It is therefore advantageous to
combine the two computations into one which is accurate at all scales.
A phenomenological way of doing so, the so-called  Santander matching, has been
proposed in Ref.~\cite{Harlander:2011aa}: it consists of 
simply interpolating between the four- and five-flavor scheme results 
by mean of a weighted average, such that 
in the two limits  $\mu/m_b \gg 1$ or $ \mu/m_b\sim 1 $ the massless or
massive results are respectively reproduced.

However, a more systematic approach which preserves the perturbative
accuracy of both computations may be desirable. One such approach,
the FONLL method,
was proposed in Ref.~\cite{Cacciari:1998it} 
in the context of hadro-production of heavy quarks, and extended to 
deep-inelastic scattering in Ref.~\cite{Forte:2010ta}.
The basic idea of this method is to expand out the
five-flavor-scheme computation in powers of the strong coupling
$\alpha_s$, and replace a finite number of terms with their
massive-scheme counterparts. The result then retains the accuracy of
both ingredients:
at the massive level,
the fixed-order accuracy corresponding 
to the number of massive orders which have been
included (FO, or fixed order), and  at the massless
level, the logarithmic accuracy of the starting five-flavor scheme
computation (NLL, or generally subleading logarithmic\footnote{We 
will consistently use the notation N$^k$LL to refer to the resummation 
of collinear logs of the heavy quark mass, i.e. by LL we mean a computation 
in which $\left(\alpha_s \ln\frac{m^2_b}{\mu^2}\right)$ is treated 
as order one ($\alpha_s^0$).}).

It is the purpose of this paper to present the application of the
FONLL scheme to  Higgs production  
in bottom-quark fusion, focusing for definiteness on the total cross-section.
In the rest of this paper we will follow the notation and conventions
of Ref.~\cite{Forte:2010ta}.

The total cross-section $\sigma$ in the 
five-flavor scheme has the form
\begin{equation}
\label{massless}
\sigma^{(5)}=\iint dx_1 dx_2\sum_{ij}f_{i}^{(5)}(x_1,\mu^2)f_j^{(5)}(x_2,\mu^2)
\,\hat{\sigma}_{ij}^{(5)}\left(x_1,x_2,\alpha_s^{(5)}(\mu^2)\right),
\end{equation}
where the sum runs over the $10$ quarks and antiquarks and the gluon,
and the $b$ quark and antiquark are treated as the other partons,
which in particular contribute to the running of
$\alpha_s^{(5)}$. For simplicity we  omit the dependence of the
hard cross-section on the renormalization and factorization scales,
which henceforth we will assume to be chosen equal to
$\mu_R=\mu_F=\mu$, unless otherwise stated.

In the four-flavor scheme it has the form
\begin{equation}
\label{massive}
\sigma^{(4)}=\iint dx_1 dx_2\sum_{ij}f_{i}^{(4)}(x_1,\mu^2)f_j^{(4)}(x_2,\mu^2)\hat{\sigma}_{ij}^{(4)}\left(x_1,x_2,\frac{\mu^2}{m_b^2},\alpha_s^{(4)}(\mu^2)\right),
\end{equation}
where now the sum only runs over the four lightest quarks and
antiquarks and the gluon, the $b$-quark decouples from the running of 
$\alpha_s^{(4)}$ and the DGLAP evolution equations satisfied by
$f_{i}^{(4)}(x_1,\mu^2)$, but full $m_b$ dependence of  the
partonic cross-section $\hat{\sigma}_{ij}^{(4)}$ is retained.

In order to carry out the FONLL procedure, we need to express the
four-flavor  scheme cross-section, Eq.~(\ref{massive}), 
in terms of $\alpha_s^{(5)}$ and $f_i^{(5)}$, so that their
perturbative expansions can be compared directly.
The coupling constant and the PDFs are related in 
the two schemes by equations of the form
\begin{equation}
\label{alphas:1}
\alpha_s^{(5)}(\mu^2)=\alpha_s^{(4)}(\mu^2)+\sum_{i=2}^\infty c_i(L)\times\left(\alpha_s^{(4)}(m^2_b)\right)^i,
\end{equation}
\begin{equation}
\label{pdf:1}
f_i^{(5)}(x,\mu^2)=\int_x^1\frac{dy}{y}\sum_{j}K_{ij}\left(y,L,\alpha_s^{(4)}(\mu^2)\right)f_j^{(4)}\left(\frac{x}{y},\mu^2\right),
\end{equation}
where
\begin{equation}
L\equiv\ln\mu^2/m_b^2
\end{equation}
and the sum runs over the eight lightest flavors, antiflavors, and the
gluon, while the index $i$ takes value over all ten quarks and
antiquarks and the gluon.
The coefficients $c_i(L)$ are polynomials in $L$, and the functions $K_{ij}$ can be expressed as an expansion in powers of $\alpha_s$, with coefficients that are polynomials in $L$.

The first nine equations~(\ref{pdf:1}) relate the eight lightest
quarks and the gluon in the two schemes and can be inverted to express
the four-flavor-scheme PDFs in terms of the five-flavor-scheme ones.  The
last two equations, assuming that the bottom quark is generated by
radiation from the gluon (i.e. no ``intrinsic''~\cite{Brodsky:1980pb}
bottom component) express the bottom and anti-bottom PDFs 
in terms of the other ones. In particular, this assumption
implies that the $b$ quark and antiquark PDFs are equal to each other, $f_b^{(5)}=f_{\bar b}^{(5)}$. 
Inverting Eqs.~(\ref{alphas:1}-\ref{pdf:1}) and substituting in Eq.~(\ref{massive}) one can obtain an expression of $\sigma^{(4)}$ in terms of $\alpha_s^{(5)}$ and $f_i^{(5)}$:
\begin{equation}
\label{massive:1}
\sigma^{(4)}=\iint dx_1 dx_2\sum_{ij=q,g}f_{i}^{(5)}(x_1,\mu^2)f_j^{(5)}(x_2,\mu^2)B_{ij}^{(4)}\left(x_1,x_2,\frac{\mu^2}{m^2_b},\alpha_s^{(5)}(\mu^2)\right),
\end{equation}
where the coefficient functions $B_{ij}$ are such that substituting the matching relations Eqs.(\ref{alphas:1})-(\ref{pdf:1}) 
in Eq.~(\ref{massive:1}) the original expression Eq.~(\ref{massive}) is recovered.  Note that in the course of the
procedure of expressing $\sigma^{(4)}$ in terms of $\alpha_s^{(5)}$
and $f_i^{(5)}$, subleading terms are introduced, because
Eqs.~(\ref{alphas:1}-\ref{pdf:1}) are only inverted to finite
perturbative accuracy. It follows that the expressions Eq.~(\ref{massive})
and  Eq.~(\ref{massive:1}) of  $\sigma^{(4)}$  actually differ by
subleading terms. Henceforth, for $\sigma^{(4)}$   we will use 
the expression  Eq.~(\ref{massive:1}), and avoid any further reference
to $\alpha_s^{(4)}$ and $f_i^{(4)}$; therefore, from now on  $\alpha_s$
and $f_i$ will denote the five-flavor scheme expressions.

In order to match the two expressions for $\sigma$ in the five-flavor
scheme, Eq.~(\ref{massless}), and in the four-flavor scheme,
Eq.~(\ref{massive:1}), we now work out their perturbative
expansion. Using DGLAP evolution, the $b$-PDF, $f_b^{(5)}(\mu^2)$, can be determined in terms of the gluon and the light-quark parton distributions $f_i^{(5)}$ at the scale $\mu^2$ convoluted with coefficient functions expressed as a power series in $\alpha_s^{(5)}$, with coefficients that are polynomials in $L$.
The five-flavor-scheme expression Eq.~(\ref{massless}) may thus be written entirely in terms of light-quark and gluon PDFs:
\begin{equation}
\label{massless:1}
\sigma^{(5)}=\iint dx_1 dx_2\sum_{ij=q,g}f_{i}^{(5)}(x_1,\mu^2)f_j^{(5)}(x_2,\mu^2)A_{ij}^{(5)}\left(x_1,x_2,L,\alpha_s^{(5)}(\mu^2)\right),
\end{equation}
where the $A_{ij}^{(5)}$ coefficient functions are given by a perturbative expansion of the form
\begin{equation}
\label{massless:coefficients}
A_{ij}^{(5)}\left(x_1,x_2,L,\alpha_s^{(5)}(\mu^2)\right)=\sum_{p=0}^N\left(\alpha_s^{(5)}(\mu^2)\right)^p\sum_{k=0}^\infty A_{ij}^{(p),(k)}(x_1,x_2)\left(\alpha_s^{(5)}(\mu^2)L\right)^k,
\end{equation}
with at leading order $N=0$, and at $\mbox{N}^m\mbox{LO}$ order $N=m$.

On the other hand, the four-flavor-scheme expression
Eq.~(\ref{massive:1}), as mentioned, is also written in terms of the 
light-quark PDFs, with coefficient functions $B_{ij}$ which can also be expanded in power of $\alpha_s^{(5)}$,
\begin{equation}
\label{massive:exp}
B_{ij}^{(4)}\left(x_1,x_2,\frac{\mu^2}{m^2_b},\alpha_s^{(5)}(\mu^2)\right)=\sum_{p=0}^N\left(\alpha_s^{(5)}(\mu^2)\right)^pB_{ij}^{(p)}\left(x_1,x_2,\frac{\mu^2}{m^2_b}\right),
\end{equation}
where $N$ is the order of the expansion needed to reach the desired
accuracy. It follows that the sum of all contributions to the
four-flavor-scheme expression Eq.~(\ref{massive:exp}) which do not
vanish when $\mu^2\gg m^2_b$ must also be present in the
five-flavor-scheme result. 

These contributions $B_{ij}^{(0),(p)}$
provide the massless limit of $B_{ij}^{(p)}$, in the sense that
\begin{equation}
\lim_{m_b\rightarrow 0}\left[B_{ij}^{(p)}\left(x_1,x_2,\frac{\mu^2}{m^2_b}\right)-B_{ij}^{(0),(p)}\left(x_1,x_2,\frac{\mu^2}{m^2_b}\right)\right]=0.
\end{equation}
In other words, $B_{ij}^{(0),(p)}$ is obtained from $B_{ij}^{(p)}$ by retaining all logarithms and 
constant terms and dropping all terms suppressed by powers of $m_b/\mu$. Given that these terms are also 
present in the five-flavor-scheme calculation, we can also write
\begin{equation}
\label{massless_lim}
B_{ij}^{(0),(p)}\left(x_1,x_2,\frac{\mu^2}{m^2_b}\right)=\sum_{k=0}^pA_{ij}^{(p-k),(k)}\left(x_1,x_2\right)L^k
\end{equation}
and
\begin{equation}
\label{massless_lim:exp}
B_{ij}^{(0)}\left(x_1,x_2,\frac{\mu^2}{m^2_b},\alpha_s^{(5)}(\mu^2)\right)=\sum_{p=0}^N\left(\alpha_s^{(5)}(\mu^2)\right)^pB_{ij}^{(0),(p)}\left(x_1,x_2,\frac{\mu^2}{m^2_b}\right).
\end{equation}
We finally define the massless limit of the four-flavor-scheme cross-section, namely
\begin{equation}
\label{massless_limit_full}
 \sigma^{(4),(0)}=\iint dx_1 dx_2\sum_{ij=q,g}f_{i}^{(5)}(x_1,\mu^2)f_j^{(5)}(x_2,\mu^2)B_{ij}^{(0)}\left(x_1,x_2,\frac{\mu^2}{m^2_b},\alpha_s^{(5)}(\mu^2)\right).
\end{equation}

The FONLL method can thus be stated as follows: replace in the five-flavor scheme expression, Eq.~(\ref{massless:1}), 
all contributions to the expansion Eq.~(\ref{massless:coefficients}) of the coefficients 
$A_{ij}^{(5)}\left(x_1,x_2,L,\alpha_s^{(5)}(\mu^2)\right)$ which appear in $B_{ij}^{(0),(p)}$, Eq.~(\ref{massless_lim}), 
with their fully massive expression $B_{ij}^{(p)}$ from Eq.~(\ref{massive:exp}). In this way, all mass suppressed effects that 
are not present in Eq.~(\ref{massless}) but are known from Eq.~(\ref{massive}), are included.
More symbolically
\begin{equation}
\label{FONLL}
\sigma^{FONLL}=\sigma^{(4)}+\sigma^{(5)}-\sigma^{(4),(0)}.
\end{equation}
If the five-flavor scheme computation is performed to N$^k$LL
accuracy, and the replacement is performed up to fixed N$^j$LO in
$\alpha_s^{(5)}$, the final result retains  N$^k$LL 
accuracy at  the massless level, and N$^j$LO accuracy at the massive
level. 

In Ref.~\cite{Forte:2010ta}, three combinations were considered
specifically in the case of deep-inelastic scattering: 
namely FONLL-A, corresponding to NLL-LO, FONLL-B,
NLL-NLO, and FONLL-C, NNLL-NLO (where by ``leading'' we always mean the first order at which the
result does not vanish, assuming no intrinsic heavy quarks). In deep-inelastic scattering, the
leading order is $O(\alpha_s^0)$ (parton model)
in the five-flavor scheme, and   $O(\alpha_s)$ in the four-flavor
scheme: there is thus a mismatch by one order, and therefore FONLL-A
is the simplest nontrivial scheme. In the case of Higgs production
in bottom fusion, the mismatch is now by two orders: the
leading order is $O(\alpha_s^0)$ (parton model)
in the five-flavor scheme, and $O(\alpha_s^2)$ in the four-flavor
scheme. 
The simplest nontrivial case, which we will also refer to as
FONLL-A, is thus NNLL-LO; we will then call FONLL-B the NNLL-NLO
combination and FONLL-C N$^3$LL-NLO.
In the five-flavor scheme, the result is known up to
NNLO~\cite{Harlander:2003ai}, thereby allowing for an NNLL computation
when used in conjunction with NNLO PDFs, and in the  
four-flavor scheme up to NLO~\cite{Dittmaier:2003ej,Dawson:2003kb},
hence in principle FONLL-A and FONLL-B are accessible using current knowledge.

\begin{figure}[ht]
\begin{center}
\begin{tabular}{ccc}
\kern -25pt
\hspace{1.5cm} \includegraphics[scale=.3]{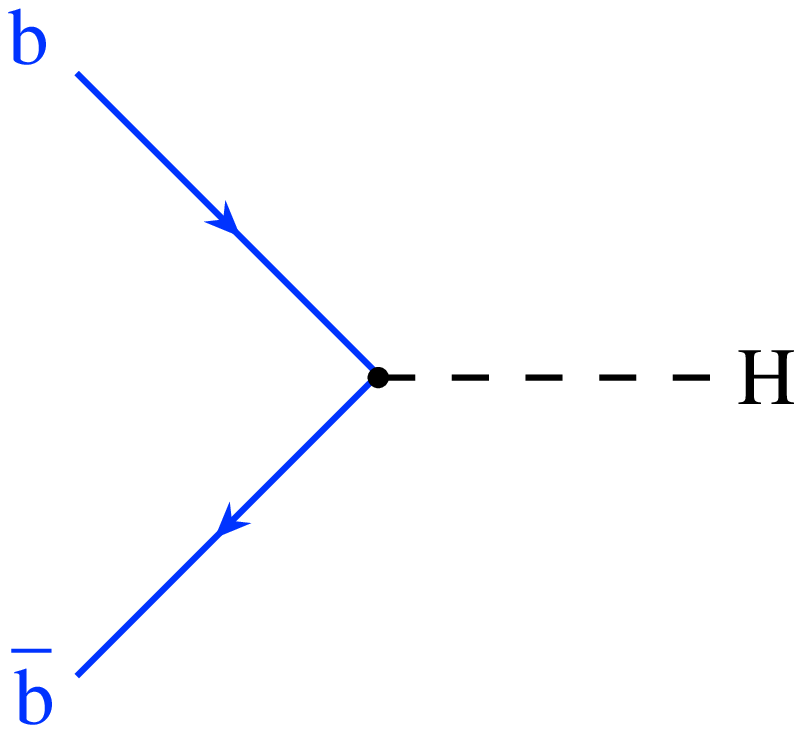} &\hspace{-2cm}
\includegraphics[scale=.3]{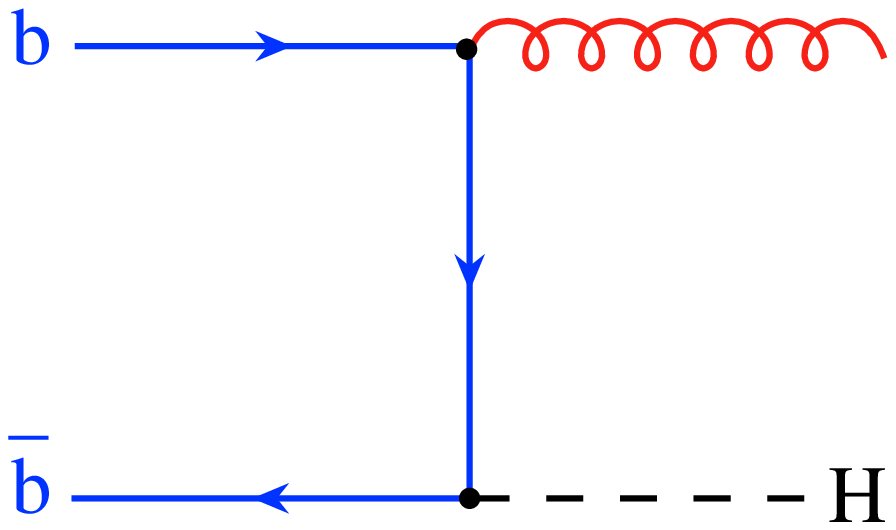} &\hspace{-2cm}
\includegraphics[scale=.3]{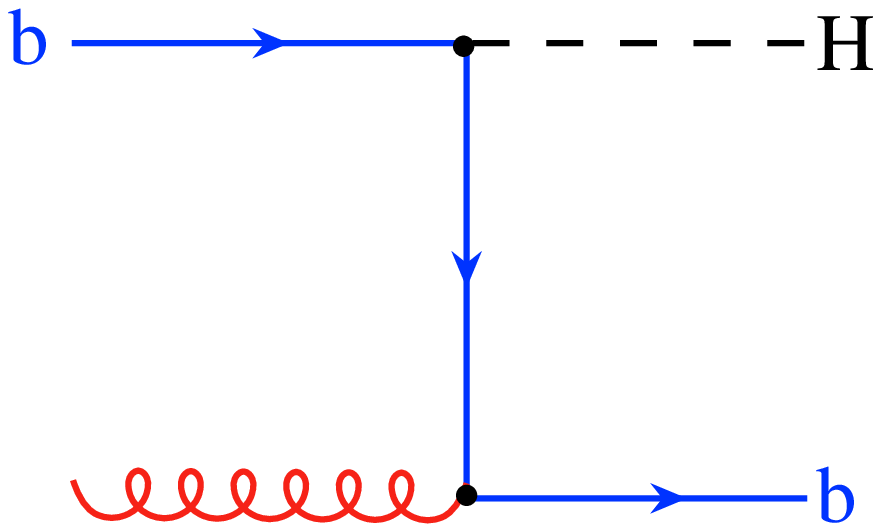}\\
\hspace{1.5cm} (a) &\hspace{-2cm} (b)&\hspace{-2cm}(c) \\[2em]
%(c) & (d) \\[2em]
\end{tabular}
\caption[]{\label{fig::bbh} Leading-order (a) and next-to-leading
  order (b-c) contributions to the hard cross-section in  the five-flavor scheme. To order $\order{\alpha_s^2}$ these processes receive 2-loop corrections (a) and 1-loop corrections (b) and (c), respectively.}
\end{center}
\end{figure}

\begin{figure}[ht]
\begin{center}
\begin{tabular}{cc}
\kern -25pt
\includegraphics[scale=.3]{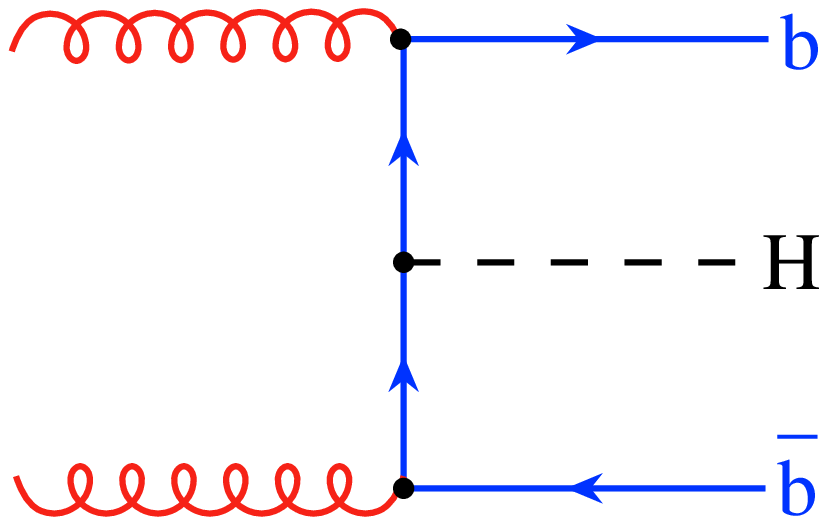} &
\includegraphics[scale=.3]{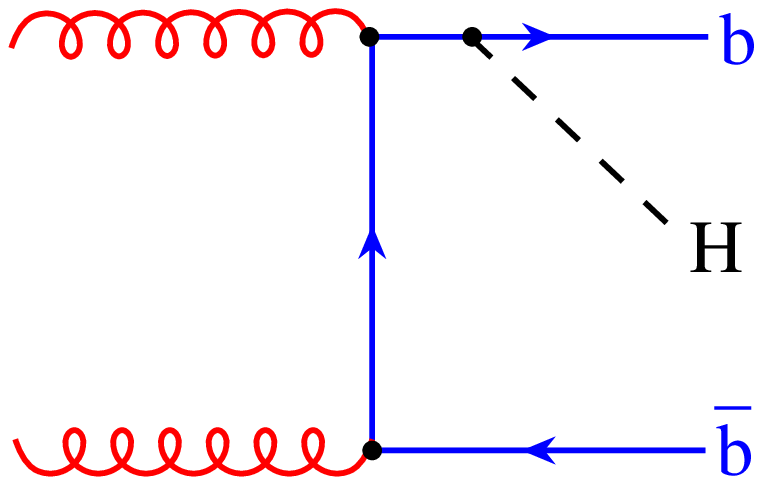} \\
%(a) & (b) \\[2em]
\kern -25pt
\includegraphics[scale=.3]{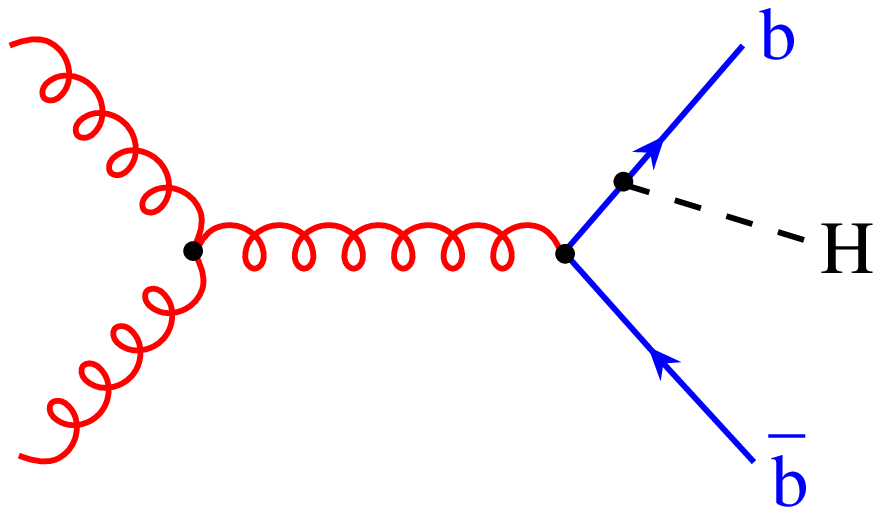} &
\includegraphics[scale=.3]{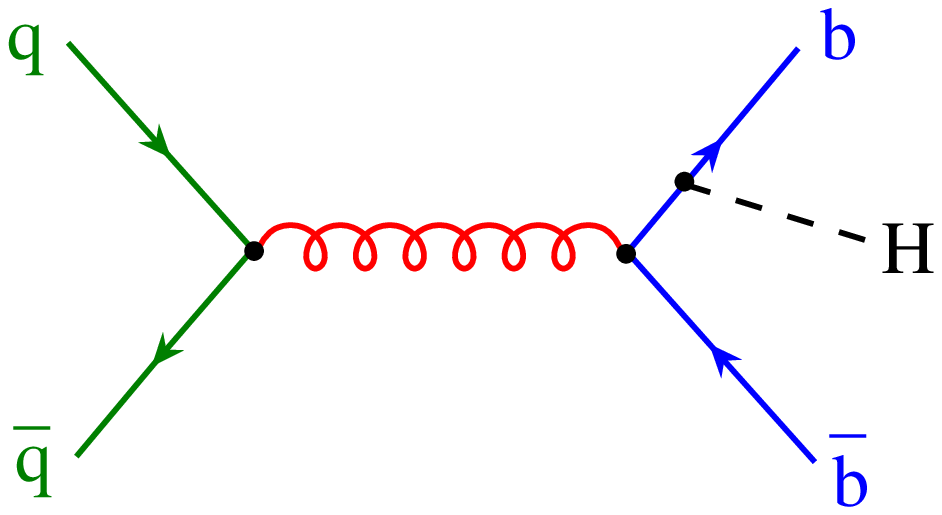} \\
%(c) & (d) \\[2em]
\end{tabular}
\caption[]{\label{fig::gghbb} Leading-order order contributions to the four-flavor scheme. Not shown are diagrams that can be obtained by crossing the initial
  state gluons, or radiating the Higgs off an anti bottom quark.}
\end{center}
\end{figure}

%%%%%%%%%%%%%%%%%%%%%%%%%%%%%%%%%%%%%
%\subsubsection{The five-flavor scheme}
We now work out Eq.~(\ref{FONLL}) explicitly  for Higgs production in
bottom-quark fusion, in the simplest FONLL-A
case.\footnote{A matched computation for the related process of Higgs
  production in top fusion has been presented
  recently~\cite{Han:2014nja}, based on a  modified version of the
  ACOT~\cite{Aivazis:1993pi} matching scheme, which for NLO
  deep-inelastic scattering is
  known~\cite{LHhq}  to coincide with FONLL-A; however, in this work
only terms up
  to NLL in the five-flavor computations are included.}
 To NNLL, the partonic cross-section must be computed up to
order $\order{\alpha_s^2}$: it then  receives contributions from 
the following sub-processes:
\begin{itemize}
\item{$\order{1} \Rightarrow b\bar{b}\rightarrow h$}
\item{$\order{\alpha_s} \Rightarrow b\bar{b}\rightarrow h$ (1-loop), $bg\rightarrow hb$, $b\bar{b}\rightarrow hg$}
\item{$\order{\alpha_s^2} \Rightarrow b\bar{b}\rightarrow h$ (2-loop), $bg\rightarrow hb$ (1-loop), $b\bar{b}\rightarrow hg$ (1-loop), $bq \rightarrow hbq$, $gg \rightarrow hb\bar{b}$, $bb \rightarrow hb\bar{b}$, $q\bar{q} \rightarrow hb\bar{b}$. }
\end{itemize}
The LO diagrams are shown in Fig.~\ref{fig::bbh}.
The full calculation up to $\order{\alpha_s^2}$ can be found in
Ref.~\cite{Harlander:2003ai}. %; results are collected in Appendix-\ref{sec:partresults} for reference.
The  relevant perturbative orders in each parton channel are thus
\begin{align}
\hat{\sigma}_{b\bar{b}}^{(5)}\left(x_1,x_2,\alpha_s^{(5)}(\mu^2)\right)&=\hat{\sigma}_{b\bar{b}}^{(5),(0)}(x_1,x_2)+\alpha_s^{(5)}(\mu^2)\hat{\sigma}_{b\bar{b}}^{(5),(1)}(x_1,x_2)\nonumber \\
&+\left(\alpha_s^{(5)}(\mu^2)\right)^2\hat{\sigma}_{b\bar{b}}^{(5),(2)}(x_1,x_2)+\order{\alpha_s^3},
\end{align}

\begin{align}
\hat{\sigma}_{bg}^{(5)}\left(x_1,x_2,\alpha_s^{(5)}(\mu^2)\right)&=\alpha_s^{(5)}(\mu^2)\hat{\sigma}_{bg}^{(5),(1)}(x_1,x_2)\nonumber \\ &+\left(\alpha_s^{(5)}(\mu^2)\right)^2\hat{\sigma}_{bg}^{(5),(2)}(x_1,x_2)+\order{\alpha_s^3},
\end{align}
\begin{equation}
\hat{\sigma}_{bq}^{(5)}\left(x_1,x_2,\alpha_s^{(5)}(\mu^2)\right)=\left(\alpha_s^{(5)}(\mu^2)\right)^2\hat{\sigma}_{bq}^{(5),(2)}(x_1,x_2)+\order{\alpha_s^3},
\end{equation}
\begin{equation}
\hat{\sigma}_{gg}^{(5)}\left(x_1,x_2,\alpha_s^{(5)}(\mu^2)\right)=\left(\alpha_s^{(5)}(\mu^2)\right)^2\hat{\sigma}_{gg}^{(5),(2)}(x_1,x_2)+\order{\alpha_s^3},
\end{equation}
\begin{equation}
\hat{\sigma}_{bb}^{(5)}\left(x_1,x_2,\alpha_s^{(5)}(\mu^2)\right)=\left(\alpha_s^{(5)}(\mu^2)\right)^2\hat{\sigma}_{bb}^{(5),(2)}(x_1,x_2)+\order{\alpha_s^3},
\end{equation}
and
\begin{equation}
\hat{\sigma}_{q\bar{q}}^{(5)}\left(x_1,x_2,\alpha_s^{(5)}(\mu^2)\right)=\left(\alpha_s^{(5)}(\mu^2)\right)^2\hat{\sigma}_{q\bar{q}}^{(5),(2)}(x_1,x_2)+\order{\alpha_s^3}.
\end{equation}

%%%%%%%%%%%%%%%%%%%%%%%%%%%%%%%%%%%%%
%\subsubsection{The four-flavor scheme}
In the four-flavor scheme, the LO $\order{\alpha_s^2}$ result corresponds 
to the  $gg\rightarrow h b\bar{b}$ and $q\bar{q}\rightarrow h b\bar{b}$  sub-processes
shown in Fig.~\ref{fig::gghbb}. 
The computation of this process in the four-flavor scheme is formally identical to that of  associate production 
of a Higgs boson with a $t \bar{t}$ pair, first performed in Ref.~\cite{Kunszt:1984ri}.

We can now match the two expressions. First, we note that in the
FONLL-A scheme the four-flavor scheme result is included to lowest
nontrivial order: therefore, we can simply replace in it
$\alpha_s^{(4)}$ and  $f_i^{(4)}$ with their five-flavor scheme
    counterparts, as the difference is higher order in $\alpha_s$  and thus
    subleading. We thus simply have
\begin{equation}
B_{ij}\left(x_1,x_2,\frac{\mu^2}{m^2_b},\alpha_s(\mu^2)\right)=\hat{\sigma}_{ij}^{(4)}\left(x_1,x_2,\frac{\mu^2}{m^2_b},\alpha_s(\mu^2)\right)+\order{\alpha_s^3}.
\end{equation}

We also need the massless limit of the four-flavor scheme
result: recalling that it starts at order $\alpha_s^2$, and using
the general expressions
Eqs.~(\ref{massless_lim})-(\ref{massless_lim:exp}), we conclude that
it must have the form
\begin{align}
B_{ij}^{(0)}(x_1,x_2,L,\alpha_s&=\left(\alpha_s\right)^2B_{ij}^{(0),(2)}\left(x_1,x_2,L\right)+\order{\alpha_s^3}\nonumber\\
&=\left(\alpha_s\right)^2\left(A_{ij}^{(2),(0)}(x_1,x_2)+A_{ij}^{(1),(1)}(x_1,x_2)L+A_{ij}^{(0),(2)}(x_1,x_2)L^2\right)
+\order{\alpha_s^3}.
\end{align}
The easiest way of determining  the coefficients 
$A_{ij}^{(p),(k)}$ is to start with the five-flavor scheme expression 
Eq.~(\ref{massless}) and expand
the bottom PDF in power of $\alpha_s$,
\begin{equation}
\label{bPDF}
f_b(x,\mu^2)=\frac{\alpha_s(\mu^2)}{2\pi} L \int_x^1 \frac{dy}{y} P_{qg}(y)f_{g}\left(\frac{x}{y},\mu^2\right)+\order{\alpha_s^2},
\end{equation}
where
\begin{equation}
P_{qg}(y)=T_R\left[y^2+(1-y)^2\right].
\end{equation}
We get
\begin{align}
A_{qq}^{(2),(0)}(x_1,x_2)&=\hat{\sigma}_{q\bar{q}}^{(5),(2)}(x_1,x_2),\\
A_{gg}^{(2),(0)}(x_1,x_2)&=\hat{\sigma}_{gg}^{(5),(2)}(x_1,x_2),\\
A_{gg}^{(1),(1)}(x_1,x_2)&=\frac{1}{2\pi}\int_0^1dy P_{qg}(y)\left(\hat{\sigma}_{gb}^{(5),(1)}(x_1,yx_2)+\hat{\sigma}_{bg}^{(5),(1)}(yx_1,x_2)\right)+(b\rightarrow\bar{b}), \\
A_{gg}^{(0),(2)}(x_1,x_2)&=\frac{1}{(2\pi)^2}\int_0^1\int_0^1dy_1 dy_2
P_{qg}(y_1)P_{qg}(y_2)\hat{\sigma}_{b\bar{b}}^{(5),(0)}(y_1x_1,y_2x_2)+(b\rightarrow\bar{b}), 
\end{align}
so that
\begin{equation}
B_{ij}^{(0),(2)}\left(x_1,x_2,L,\alpha_s\right)=A_{ij}^{(2),(0)}(x_1,x_2)+A_{ij}^{(1),(1)}(x_1,x_2)L+A_{ij}^{(0),(2)}(x_1,x_2)L^2.
\end{equation}

We now have all the ingredients which enter the FONLL-A
expression. For book-keeping purposes, we
introduce a formal expansion of the cross-section of the form
\begin{equation}
\label{FONLL:exp}
\sigma^{\rm FONLL-A}=\sigma^{\rm FONLL-A,(0)}+\alpha_s(\mu^2)\sigma^{\rm FONLL-A,(1)}+\left(\alpha_s(\mu^2)\right)^2\sigma^{\rm FONLL-A,(2)}+\order{\alpha_s^3},
\end{equation}
where it is understood that only the coefficient functions
$B_{ij}^{(4)}$, $A_{ij}^{(5)}$  and $B_{ij}^{(0)}$ in
Eqs.~(\ref{massive:1}), (\ref{massless:1}) and
(\ref{massless_lim:exp}) respectively are expanded, but not the
PDFs. The expansion is formal in that, as we have just seen,
the nominally $\order{\alpha_s^0}$ contribution really starts at
$\order{\alpha_s^2}$ once one substitutes the explicit expression
Eq.~(\ref{bPDF}) of the
$b$-quark distribution, as it should be in order for it to match the
four-flavor scheme expression.

Be that as it may, 
since the four-flavor scheme starts at  $\order{\alpha_s^2}$,
$\sigma^{\rm FONLL-A,\{(0),(1)\}}$, the first two terms in the
expansion Eq.~(\ref{FONLL:exp}) coincide with the    five-flavor scheme expressions:
\begin{equation}
\sigma^{\rm FONLL-A,(0)}=\iint dx_1 dx_2 f_{b}(x_1,\mu^2)f_{\bar{b}}(x_2,\mu^2)\hat{\sigma}_{b\bar{b}}^{(5),(0)}\left(x_1,x_2\right)
\end{equation}
\begin{align}
\sigma^{\rm FONLL-A,(1)}&=\iint dx_1 dx_2 \left\{f_{b}(x_1,\mu^2)f_{\bar{b}}(x_2,\mu^2)\hat{\sigma}_{b\bar{b}}^{(5),(1)}\left(x_1,x_2\right)\right. \nonumber \\
&\left. + \hat{\sigma}_{gb}^{(5),(1)}(x_1,x_2)\left[\left(f_g(x_1,\mu^2)f_{b}(x_2,\mu^2) +(x_1\rightarrow x_2)\right)+(b\rightarrow \bar{b})\right]\right\}.
\end{align}

The $\order{\alpha_s^2}$ contribution can be written as the sum of two
terms: four-flavor scheme, and  difference between 
the five-flavor and the massless limit of the four-flavor scheme. The
former is simply given by the leading-order partonic cross-section  
in the four-flavor scheme.The latter is given by
\begin{equation}
\sigma^{\rm FONLL-A,(2)}=\sigma^{(4),(2)}+\sigma^{(d),(2)},
\end{equation}
where
\begin{equation}
\sigma^{(d),(2)}=\sigma^{(5),(2)}-\sigma^{(4),(0),(2)},
\end{equation}
and
\begin{equation}
\sigma^{(4),(0),(2)}=\iint dx_1dx_2 \sum_{ij=q,g}f_{i}(x_1,\mu^2)f_j(x_2,\mu^2)B_{ij}^{(0),(2)}\left(x_1,x_2,L,\alpha_s\right).
\end{equation}
We get
\begin{align}
\label{sigma:diff}
\sigma^{(d),(2)}&=\iint dx_1 dx_2 \left\{\vphantom{\sum_{k}^P}f_b(x_1,\mu^2)f_{\bar{b}}(x_2,\mu^2)\hat{\sigma}_{b\bar{b}}^{(5),(2)}(x_1,x_2)\quad + \right. \nonumber \\
&+f_b(x_1,\mu^2)f_{b}(x_2,\mu^2)\hat{\sigma}_{bb}^{(5),(2)}(x_1,x_2)\quad + \nonumber \\
&+  \hat{\sigma}_{gb}^{(5),(2)}(x_1,x_2)\left[\left(f_g(x_1,\mu^2)f_{b}(x_2,\mu^2) +(x_1\rightarrow x_2)\right)+(b\rightarrow \bar{b})\right]\quad +\nonumber \\
&+ \hat{\sigma}_{qb}^{(5),(2)}(x_1,x_2)\left[\left(f_q(x_1,\mu^2)f_{b}(x_2,\mu^2) +(x_1\rightarrow x_2)\right)+(b\rightarrow \bar{b},q\rightarrow \bar{q})\right]\quad +\nonumber \\
&-\frac{L}{2\pi}\iint dyP_{qg}(y)\left[\vphantom{\sum_k}\hat{\sigma}_{bg}^{(5),(1)}(x_1,yx_2)\,f_g(x_1,\mu^2)f_{g}(x_2,\mu^2)\right.\nonumber \\ 
&\left.\qquad \qquad+ \hat{\sigma}_{bg}^{(5),(1)}(yx_1,x_2)f_g(x_1,\mu^2)f_{g}(x_2,\mu^2)+(b\rightarrow\bar{b})\vphantom{\sum_k}\right] \quad+\nonumber \\
&-\left.\frac{L^2}{4\pi^2}\iint dy_1 dy_2 P_{qg}(y_1)P_{qg}(y_2)f_g(x_1,\mu^2)f_{g}(x_2,\mu^2)\hat{\sigma}_{b\bar{b}}^{(5),(0)}(y_1x_1,y_2x_2)\vphantom{\sum_{k}^P}\right\},
\end{align}
which is our main result. Note that in the general case in which
$\mu_R\not=\mu_F$, the expansion Eq.~(\ref{FONLL:exp}) should be
viewed as an expansion in powers of $\alpha_s(\mu_R)$; the log is
$L\equiv\ln\frac{\mu_F^2}{m_b^2}$; all PDF should be evaluated at
$\mu=\mu_F$, and all five-flavor scheme partonic cross-sections
should be evaluated at the appropriate scale $\hat\sigma_{ij}^{(5),(i)}=
\hat\sigma_{ij}^{(5),(i)}(\mu^2_R,\mu^2_F)$. Strictly speaking, in
this case the argument of the strong coupling in the 
term in Eq.~(\ref{sigma:diff}) which is linear in $L$ should be
$\alpha_s(\mu_R)\alpha_s(\mu_F)=(\alpha_s(\mu_R))^2(1+O(\alpha_s^3))$.

It is easy to see explicitly that, if the $b$-PDF is  expressed in terms of
its values at $\mu^2=m_b^2$ using Eq.~(\ref{bPDF}),  the FONLL-A expression differs from
the four-flavor scheme result by terms of order $\alpha_s^3$, namely, 
the difference term
\begin{equation}\label{eq:fulldiff}
\sigma^{(d)}= \sigma^{(5),(0)} + \alpha_s(\mu^2) \sigma^{(5),(1)}+ (\alpha_s(\mu^2))^2  \sigma^{(d),(2)}
\end{equation}
is $\order{\alpha_s^3}$. Indeed, Eq.~(\ref{bPDF}) implies that all
contributions to $\sigma^{(d),(2)}$ but the logarithmic ones are
$\order{\alpha_s^3}$. We then have
\begin{align}
  \sigma^{(d)} & = \iint dx_1 dx_2 \Bigg\{\Big[ f_{b}(x_1,\mu^2)f_{\bar{b}}(x_2,\mu^2)\hat{\sigma}_{b\bar{b}}^{(5),(0)}\left(x_1,x_2\right)  \nonumber \\
    &    -\frac{\alpha_s^2L^2}{4\pi^2}\iint dy_1 dy_2 P_{qg}(y_1)P_{qg}(y_2)f_g(x_1,\mu^2)f_{g}(x_2,\mu^2)\hat{\sigma}_{b\bar{b}}^{(5),(0)}(y_1x_1,y_2x_2)\Big]\label{line:1} \\
  &  +\Big[\alpha_s\hat{\sigma}_{gb}^{(5),(1)}(x_1,x_2)\left(f_g(x_1,\mu^2)f_{b}(x_2,\mu^2) +(x_1\rightarrow x_2)\right)   \nonumber \\
    & -
    \frac{\alpha_s^2L}{2\pi}\int d yP_{qg}(y)\left(\hat{\sigma}_{bg}^{(5),(1)}(x_1,yx_2)\,+
    \hat{\sigma}_{bg}^{(5),(1)}(yx_1,x_2)\right)f_g(x_1,\mu^2)f_{g}
     (x_2,\mu^2)\Big]\nonumber \Bigg\} \,+\,\order{\alpha_s^3}. \nonumber
\end{align}
Substituting  Eq.~(\ref{bPDF}) in Eq.~(\ref{line:1}) 
all terms in Eq.~(\ref{line:1}) cancel, as expected.

We can now study the phenomenological implications of our results.
Leading-order four-flavor scheme  predictions have been
obtained using a modified version of the SHERPA Monte Carlo 
generator~\cite{Gleisberg:2008ta} which we tested against results
obtained in Ref.~\cite{Dittmaier:2003ej} and
Ref.~\cite{Wiesemann:2014ioa}; for NLO  results
(which we will also show for comparison) this has been further
interfaced to the
{\tt OpenLoops} code~\cite{Cascioli:2011va}. Four-flavor scheme results are
obtained using  $n_f=4$ {\sc NNPDF3.0} LO PDFs~\cite{Ball:2014uwa}
with $\alpha^{\rm 5F}_S(m_Z)=0.118$. Five-flavor  scheme predictions
are obtained using  the
 {\tt bbh@nnlo} code~\cite{Harlander:2003ai} with 
the $n_f=5$ NNLO {\sc NNPDF3.0} parton set~\cite{Ball:2014uwa}.
For FONLL-A, results for the central scale choice have been
obtained in two different ways. First, we have recomputed the 
four-flavor scheme result, but now using $n_f=5$ NNLO {\sc NNPDF3.0}
PDFs, and we have combined this with our implementation
of Eq.~(\ref{sigma:diff}). Then, 
we have checked that we get the same answer by combining %
this four-flavor scheme result with the five-flavor scheme one
from the {\tt bbh@nnlo} code, and adding 
an implementation of the subtraction term
Eq.~(\ref{massless_limit_full}). Scale variation plots have then been
produced using this second combination.
In all cases, the strong coupling provided
with the PDF set has been used  through the  {\sc LHAPDF}
interface~\cite{Buckley:2014ana}. The $b$ mass in FONLL
expressions has been
identified with the pole mass, for which we have taken the value
$m_b=4.72$~GeV; this  
corresponds to the $\overline{\rm MS}$ value
$\overline{m}_b(\overline{m}_b)=4.21$~GeV through
the two-loop  relation of Ref.~\cite{Kuhn:2007vp}, which we
implemented in order to 
evaluate the bottom Yukawa coupling in the $\overline{\text{MS}}$
scheme at $\mu=\mu_R$. Like $\alpha_s$ and the PDFs, 
Yukawa couplings are evolved at NNLO in the five-flavor scheme in all
contributions to the FONLL expression.

\begin{figure}[!ht]
\begin{center}
\includegraphics[width=0.7\textwidth]{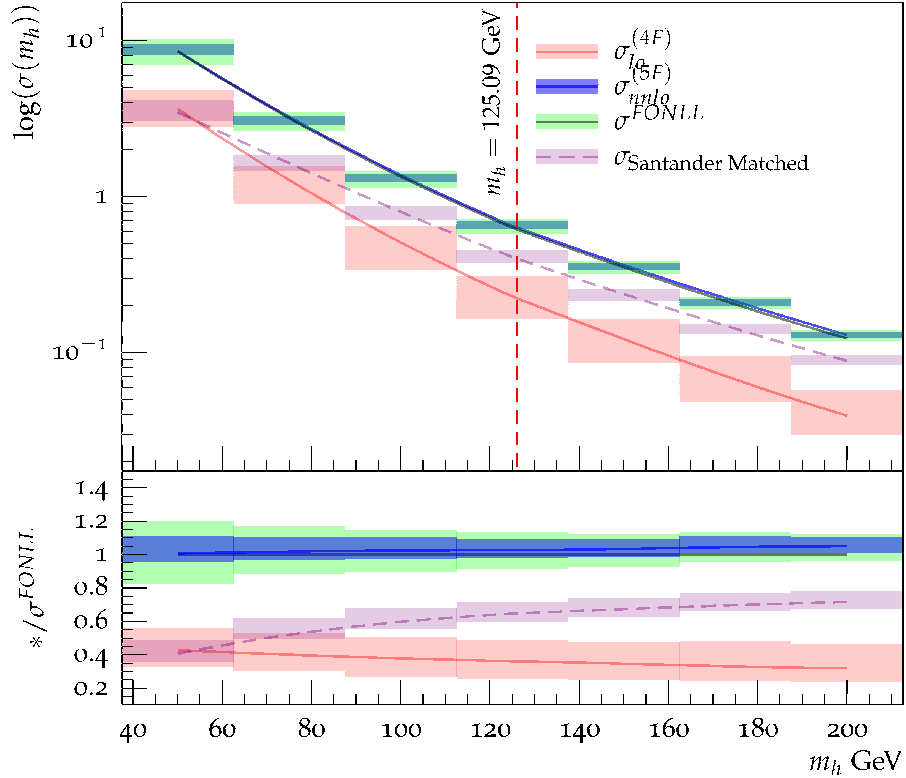} 
\caption{\label{fig:resMH} The total inclusive cross-section computed in the
  four-flavor scheme at LO (red), in the five-flavor scheme at NNLO
  (blue), and in the FONLL-A scheme (green). The Santander matching
  Eq.~\ref{eq:S-M} of the four and five-flavor scheme results is
  also shown (purple). Both the absolute result (top) and the ration
  to the FONLL-A prediction (bottom) are shown.}
\end{center}
\end{figure}

In Fig.~\ref{fig:resMH} we compare the cross-section computed in the four-flavor,
five-flavor and FONLL-A scheme. Results are shown as a function of the
Higgs mass. Here and henceforth, uncertainty 
bands are obtained by varying the renormalization
and factorization scales $\mu_R$ and $\mu_F$ independently by a factor of
2 about the central value $\mu_F=\mu_R=m_H$, discarding the two
extreme points $\mu_R= 4 \mu_F$ and $\mu_F= 4 \mu_R$, and taking the
envelope of results. In the same figure we also show the curve
obtained using the so-called Santander matching of
Ref.~\cite{Harlander:2011aa}, which is given by
\begin{equation}
\label{eq:S-M}
\sigma_{S-M} = \frac{\sigma^{(4F)}+w\,\sigma^{(5F)}}{1+w}\,.
\end{equation}
with $w = \ln{m_H/m_b}-2$: this reproduces the five-flavor scheme
result when $w\to\infty$, and the four-flavor scheme one when
$w=0$. This prescription was suggested in Ref.~\cite{Harlander:2011aa}
to be used with the highest-order available four- and five-flavor scheme
results. Here, we show it using the LO four-flavor scheme result in
order to provide a meaningful assessment of the differences in
comparison to FONLL-A.

The four-flavor scheme result is rather smaller than the five-flavor
scheme one, and it is affected by a significantly larger scale
uncertainty, as one expects of a LO computation.  
 The
FONLL and five-flavor scheme results are very
close, with, for $m_h=125.09$ GeV, 
the FONLL prediction just below the five-flavor one, with a 
somewhat larger uncertainty. Note that the four-flavor scheme result
shown in the  plot is determined using LO PDFs, while the four-flavor
scheme result that enters the FONLL combination is consistently
computed with NNLO PDFs, as discussed above. We have verified that the
latter would be yet lower, further away from the five-flavor scheme
results, as one expects due to the fact the the LO gluon is typically larger.
This shows  that mass effects for this process are  small, though 
not negligible in comparison to  the scale uncertainty on the
five-flavor result, as we will see shortly. 
The fact that mass-corrections at leading order are
small was already noticed in  Ref.~\cite{Buttar:2006zd}.
Such a quantitative conclusion cannot be arrived at using the
Santander-matched result, which simply interpolates between the four-
and five-flavor scheme results. 

\begin{figure}[!htb]
\begin{center}
\includegraphics[width=0.5\textwidth,angle=270]{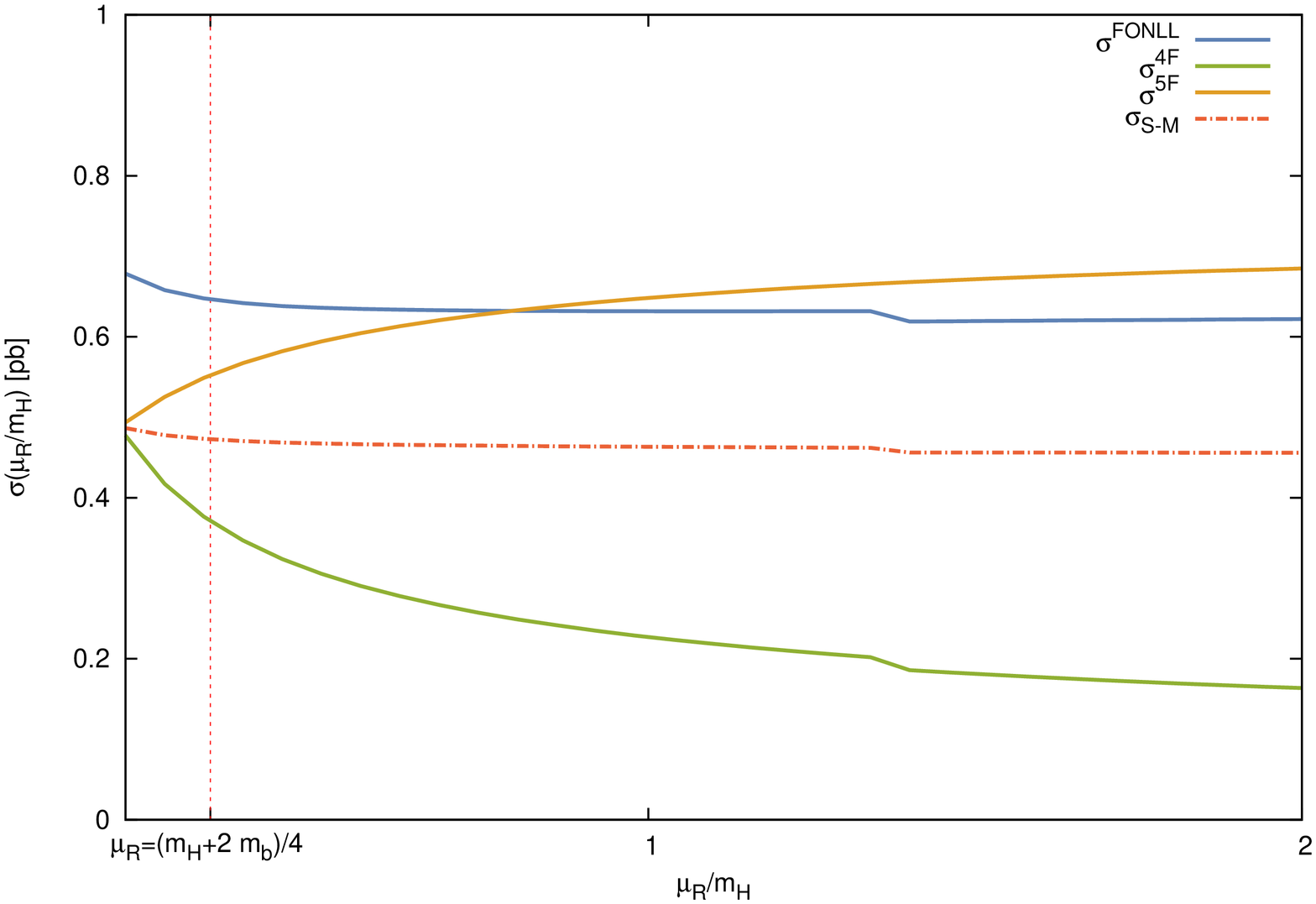} 
\includegraphics[width=0.5\textwidth,angle=270]{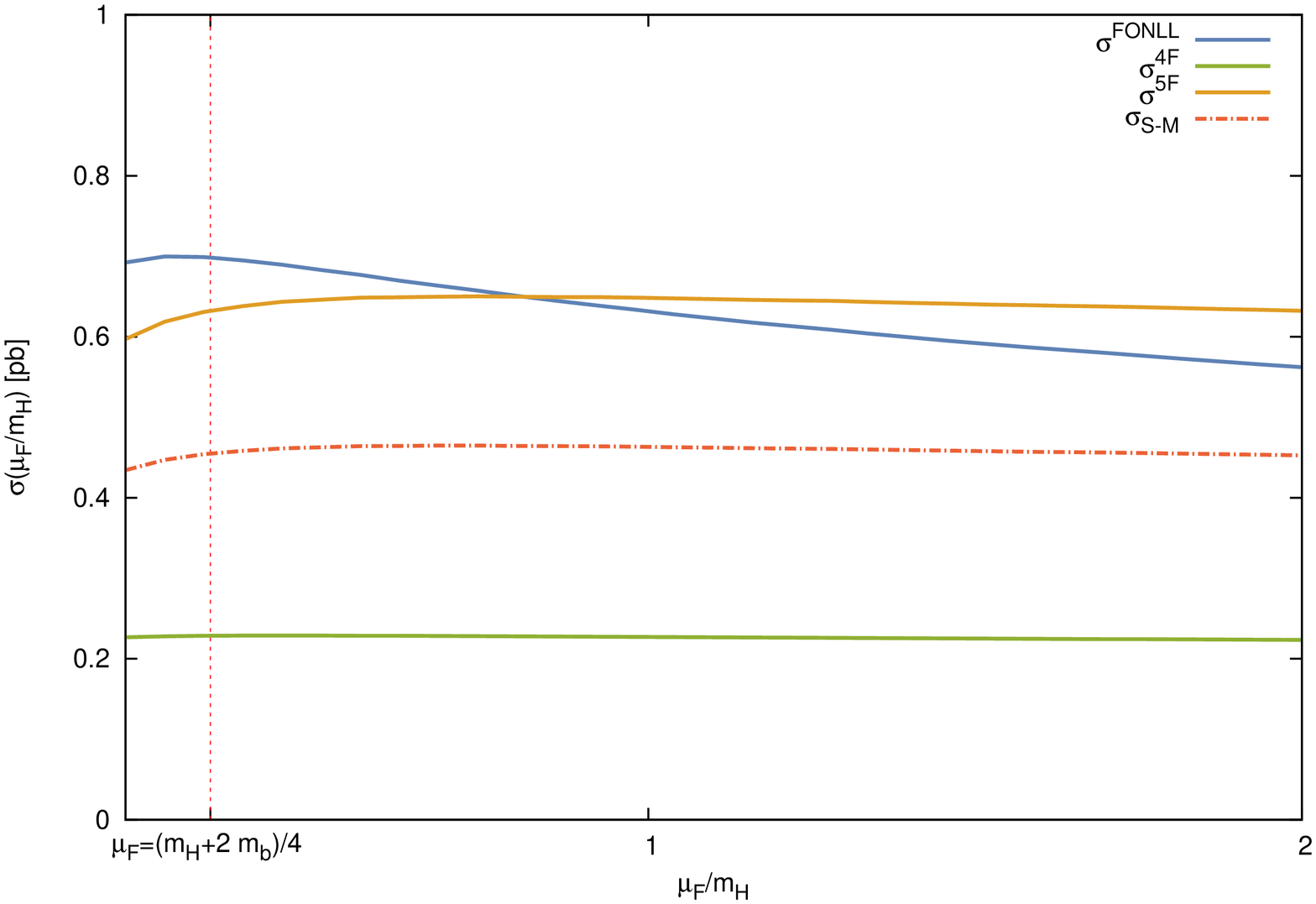} 
\caption{\label{fig:resMH4} Renormalization (top) and factorization
  (bottom) scale dependence of the cross-sections
  shown in Fig.~\ref{fig:resMH} with $m_H=125.09$~GeV. The preferred
  scale choice $\frac{m_H+2 m_b}{4}$ is denoted by a vertical bar.}
\end{center}
\end{figure}

The scale dependence of the various results of Fig.~\ref{fig:resMH} is
shown in Fig.~\ref{fig:resMH4} for $m_H=125.09$~GeV. 
The
four- and five-flavor scheme results display a significant
renormalization scale dependence. The four-flavor scheme result drops
significantly as the scale is increased because of the reduction in value of
$\alpha_s$, while the five-flavor scheme results grows  because the
residual, weaker $O(\alpha_s^3)$ dependence has the opposite sign
(NNLO corrections are negative) combines with the growth of the Yukawa
coupling with scale. Interestingly, this scale dependence  cancels to
a large extent
both in the FONLL-A and
Santander matched results. As a consequence, the mass-corrections
included in the FONLL-A result, and
the scale dependence of the five-flavor scheme computation are of
comparable size, with the FONLL result below the massless one at the
upper range  of the scale variation, and above it  for lower scale
choices, and specifically if the  renormalization scale is fixed at
 $\mu_R=\frac{m_H+2 m_b}{4}$, as recommended in
Refs.~\cite{Dittmaier:2003ej,Maltoni:2003pn,Dittmaier:2011ti}, with a
crossing point just below $\mu_R=m_H$.

The factorization scale dependence is very
mild in all schemes, except for FONLL, where it turns out that the
scale dependence is of the same order as the mass-corrections, which
as we have seen are small but not negligible. In fact, the
factorization scheme dependence shown in the plot has been determined
using as   argument of the strong coupling for  the 
term in Eq.~(\ref{sigma:diff}) which is linear in $L$ 
$\alpha_s(\mu_R)\alpha(\mu_F)$, as discussed above. If one makes the choice
$(\alpha_s(\mu_R)^2)$, which is equivalent up to subleading term, the
scale dependence changes (and in fact it becomes stronger) by an
amount which is comparable to the scale variation itself. This means
that corrections of relative order $(\alpha_s(\mu_R)^2)
\ln(\mu_R/\mu_F)$ to the mass-corrections are not negligible on the
scale of the mass-corrections themselves. They could only be accounted
for by upgrading the four-flavor scheme computation to NLO.

Finally, in Table~\ref{tab:resMH} we collect our results with
$m_H=125.09$~GeV and $\mu=m_H$ or $\mu=\frac{m_H+2 m_b}{4}$. For
comparison, in addition to the 
results shown in Figs.~\ref{fig:resMH}-\ref{fig:resMH4}
we 
also show the best available calculation in the four-flavor scheme
(NLO) 
and its Santander matching
to the NNLO five-flavor result.
 \begin{table}[!htb]
 \begin{center}
 \begin{tabular}{ccccc|cc}
 \hline \hline
      & $\sigma^{\rm (5F)}$ (pb) & $\sigma_{\rm LO}^{\rm (4F)}$ (pb) &  $\sigma^{\rm FONLL}$ (pb) &  $\sigma_A^{\rm S-M}$ (pb)& $\sigma^{\rm (4F)}$ (pb) &$\sigma^{\rm S-M}$ \\
\hline
 $\mu=m_H$ & $0.65_{-0.03}^{+0.07} $ &   $ 0.22^{+0.25}_{-0.06}$& $0.63_{-0.01}^{+0.34}$   & $ 0.55^{+0.20}_{-0.10}$   & $0.26_{-0.10}^{+0.19}$  &  $0.56_{-0.13}^{+0.12} $ \\
 &                   &   &      &             &  &   \\
 $\mu=(m_H+2 m_b)/4$ &  $0.61$  &  $ 0.41 $ & $ 0.82$    &  $0.56 $ & $ 0.42$ & $0.57 $ \\
 \hline \hline
 \end{tabular}
 \end{center}
 \caption{\label{tab:resMH} The total cross-section computed for
   $m_H=125.09$~GeV in the five-flavor scheme at NNLO, the four-flavor
   scheme at LO, and matching the two with FONLL-A, or with Santander
   matching (denoted as $\sigma_A^{\rm S-M}$). The NLO four-flavor
   scheme result, and its Santander matching to the five-flavor scheme
   are also shown for comparisons. Results are given for  $\mu=m_H$
   (top row) and  $\mu=(m_H+2 m_b)/4$ (bottom row). For $\mu=m_H$ we
   also show the uncertainty band obtained from scale variation (see text).}
 \end{table}

In summary, we have shown how to consistently match the four- and
five-flavor scheme computations of Higgs production in bottom-quark
fusion. We have found that a fully matched computation allows detailed
quantitative comparisons between the computations in various schemes,
unlike other more phenomenological approaches. However,
for competitive precision phenomenology, the results presented in
this paper should be upgraded to include the four-flavor scheme result
up to NLO: indeed, the factorization scheme dependence of the
mass corrections turns out to be comparable to their size. 
Such an upgrade is possible  by using the scheme
presented here, in its FONLL-B version, which requires an in principle
straightforward, though in practice somewhat laborious extension of
the techniques presented in this paper: this is the object of ongoing work.

\section*{Acknowledgments}
We thank Fabio Maltoni for several illuminating discussions. 
We thank Marius Wiesemann for his help in comparing our results 
to those obtained with MG5.
SF and DN are  supported by the European Commission through the
HiggsTools Initial Training Network  PITN-GA2012-316704,  SF also
by an Italian PRIN2010 grant, and 
MU  by the UK Science and Technology Facilities Council.
%%%%%%%%5
%\begin{appendix}
%\include{app}
%\end{appendix}

%\renewcommand{\em}{}
%\bibliographystyle{UTPstyle}
%\bibliography{bbH_FONLL}
\providecommand{\href}[2]{#2}\begingroup\raggedright\endgroup

\end{document}